\documentclass[a4paper,11pt]{article}
\usepackage{pos}

\newcommand{\znubb}{$0\nu\beta\beta$ }
\newcommand{\znubbns}{$0\nu\beta\beta$}
\newcommand{\thalf}{T_{1/2}^{0\nu}}
\newcommand{\amp}{\mathcal{A}}
\newcommand{\M}{\overline{M}}

\renewcommand{\logo}{\relax}

\title{Neutrinoless double beta decay rates and the $3+2$ scenario}
\ShortTitle{\znubb and 2 HNLs}

\author*[a,b]{Vaisakh Plakkot}

\affiliation[a]{Institute of Physics, University of Amsterdam,\\
  Science Park 904, 1098 XH Amsterdam, The Netherlands}

\affiliation[b]{Theory Group, Nikhef,\\
Science Park 105, 1098 XG Amsterdam, The Netherlands}

\emailAdd{v.plakkot@uva.nl}

\abstract{
The possible Majorana nature of neutrinos leads to lepton-number-violating effects such as neutrinoless double beta decay. The standard study of this process involves mass-dependent matrix elements which, although easy to use, might be missing important effects, especially in the light neutrino regime where the ultrasoft contributions become important. A fresh look at the different momentum regions leads us to an effective practical parametrisation for the decay amplitude that can show significant differences in the light-to-medium mass range of neutrinos compared to the standard parametrisation. As a concrete realisation of a UV model leading to Majorana neutrinos, the testability of a $3+2$ model with two sterile neutrinos is discussed.}

\FullConference{The 11th International Workshop on Chiral Dynamics (CD2024)\\
 26-30 August 2024\\
Ruhr University Bochum, Germany\\}

\begin{document}
\maketitle

\section{Introduction}

Neutrino oscillations~\cite{Super-Kamiokande:1998kpq}, pointing towards non-zero Standard Model (SM) neutrino masses, is an outstanding problem with its solution perhaps lying somewhere in the realm of Beyond-the-Standard-Model (BSM) physics. Neutrino masses can be introduced at mass dimension-five in Effective Field Theory (EFT) in the form of the Weinberg operator, giving neutrinos a Majorana nature. This would then point towards the possibility of lepton-number-violating (LNV) effects, such as neutrinoless double beta decay (\znubbns). Given the rapid experimental progress, with the possibility to probe the inverted mass ordering band for three light Majorana neutrinos in the next few decades~\cite{LEGEND:2017cdu,nEXO:2017nam}, it is imperative to revisit the calculation of these decay rates to equip ourselves with more accurate predictions when the time comes. To this end, recent computations provide improvements to the standard rate calculations which can miss important contributions to the neutrino exchange amplitude~\cite{Cirigliano:2018hja,Dekens:2023iyc,Dekens:2024hlz}.

From a high-energy perspective, the Weinberg term can be obtained by integrating out massive (Majorana) neutrino degrees of freedom, or heavy neutral leptons (HNLs), sometimes also called sterile neutrinos -- typically obtained through the addition of right-handed neutrinos (RHNs) to the SM, and the seesaw mechanism (see, e.g., refs.~\cite{Drewes:2013gca,King:2025eqv} for reviews). These HNLs can leave signatures in collider and fixed target experiments, or cosmological and astrophysical events, or both, in addition to contributing to \znubb (see, e.g., refs.~\cite{Bolton:2019pcu,Bolton:2022tds}). HNLs can also lead us towards the revelation to the mystery of matter-antimatter asymmetry via leptogenesis. We show how a minimal BSM scenario with two HNLs can be completely tested in the near future for the inverted mass ordering of neutrinos (IO), with the help of \znubb and collider experiments, as well as the requirement to produce the correct amount of baryon asymmetry in the universe (BAU).

\section{Neutrinoless double beta decay rates}

The half-life of an isotope that can undergo \znubb is given as
\begin{align}\label{LifetimeFormula}
    \left(\thalf\right)^{-1} = G_{01}\, g_A^4 \left| \sum_{i} V_{ud}^2\, \frac{m_i}{m_e}\,\mathcal{U}_{ei}^2\,\amp(m_i)\right|^2,
\end{align}
where $G_{01}$ is an isotope-dependent phase-space factor, $g_A \simeq 1.27$ is the nucleon axial coupling, $V_{ud}\simeq 0.97$ is the up-down CKM matrix element, $m_e$ is the electron mass, and $i$ runs over \emph{all} neutrino mass eigenstates. The mixing of mass eigenstates with the electron neutrino flavour is denoted with $\mathcal{U}_{ei}$, and the mass-dependent amplitude $\amp(m_i)$ captures all hadronic and nuclear physics. The rest of this section will focus on the calculation of these amplitudes.

\subsection{Standard computation}

For very small neutrino masses, $m_i \ll m_\pi$, the amplitude saturates and one can effectively set the mass to zero for the purpose of calculating the amplitudes. $\amp(0)$ contains a short-distance and a long-distance component, and is given in terms of the Fermi, Gamow-Teller, and tensor matrix elements ($\mathcal{M}_F,\,\mathcal{M}_{GT},\,\mathcal{M}_{T}$), QCD matrix element $g_\nu^{NN}$, and the short-distance nuclear matrix element (NME) $\mathcal{M}_{F,\text{sd}}$ as
\begin{align}
\amp(0) = -\frac{\mathcal{M}_F}{g_A^2} + \mathcal{M}_{GT} + \mathcal{M}_T - 2 g_\nu^{NN} m_\pi^2 \frac{\mathcal{M}_{F,sd}}{g_A^2}\,.
\end{align}

For heavier neutrinos, for instance in case of HNLs, the neutrino propagator is effectively modified to include masses, i.e., $\frac{1}{k_i^2}\rightarrow \frac{1}{k_i^2 - m_i^2}$, and the mass-dependent amplitude is usually assumed to take the functional form
\begin{align}
\amp(m_i) = -\mathcal{M}(0) \frac{\langle p^2 \rangle}{\langle p^2 \rangle + m_i^2}\,,
\label{eq:stdformula}
\end{align}
where $\langle p^2 \rangle\sim m_\pi^2$ is a nucleus-dependent fit parameter, and the NME $\mathcal{M}(0)$ is computed within nuclear many-body frameworks. Although simple to use and effective in many ways, this formula misses several potentially important contributions.

\subsection{Updated computation}

The neutrino momenta in \znubb (denoted with ``$k$'' below) can be divided into separate regions:
\begin{itemize}
\item Hard region, with $k_0 \sim |\vec{k}| \sim \Lambda_\chi$,
\item Soft region, with $k_0 \sim |\vec{k}| \sim m_\pi$,
\item Potential region, with $k_0 \sim \frac{|\vec{k}|^2}{m_N} \sim \frac{m_\pi^2}{m_N}$,
\item Ultrasoft region, with $k_0 \sim |\vec{k}| \sim \frac{m_\pi^2}{m_N}$,
\end{itemize}
where $\Lambda_\chi \sim$ GeV is the chiral scale, and $m_\pi$ and $m_N$ are the pion and nucleon masses. The parametrisation in eq.~\eqref{eq:stdformula} typically only includes contributions from the potential region.

From an EFT perspective, one should consider different mass regions of neutrinos separately, derive the amplitude for each region using the relevant EFT, and match them at the boundaries. We thus use the following division:
\begin{align}
\amp(m_i) = 
\begin{cases}
\amp^\text{(p,<)}(m_i) + \amp^\text{(h)}(m_i) + \amp^\text{(us)}(m_i) &\text{ for} \quad m_i<100\text{ MeV}\,,\\
\amp^\text{(p)}(m_i) + \amp^\text{(h)}(m_i) &\text{ for} \quad 100\text{ Mev }\leq m_i<2\text{ GeV}\,,\\
\amp^{(9)}(m_i) &\text{ for} \quad 2\text{ Gev }\leq m_i\,.
\end{cases}
\end{align}
At large masses $(\geq 2 \text{ GeV})$, the heavy neutrino is integrated out at the quark level, leaving us with a contact interaction at dim-9, giving the amplitude $\amp^{(9)}$. For intermediate masses, the hard and potential regions are most relevant, giving us the contributions $\amp^\text{(p)}$ and $\amp^\text{(h)}$ respectively. For smaller masses, the ultrasoft region enters the game, and one must now also account for double-counting of the linear terms in the potential region, giving us the corrected piece $\amp^{(p,<)}$ on top of the ultrasoft contribution $\amp^\text{(us)}$. The exact form of the amplitudes, and the values of parameters involved, can be found in ref.~\cite{Dekens:2024hlz}.

\begin{figure}[h!]
\centering
\begin{minipage}[t]{.48\textwidth}
\centering
\includegraphics[width=\linewidth]{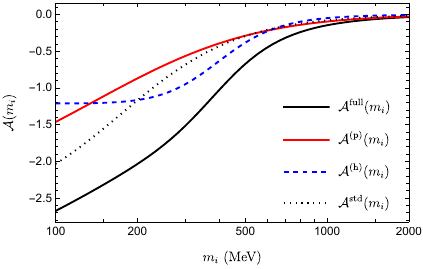}
\caption{Contribution to the \znubb amplitude from a single neutrino with mass $m_i$. Figure adapted from ref.~\cite{Dekens:2024hlz}.}
\label{fig:amps}
\end{minipage}
\hfill
\begin{minipage}[t]{.48\textwidth}
\centering
\includegraphics[width=\linewidth]{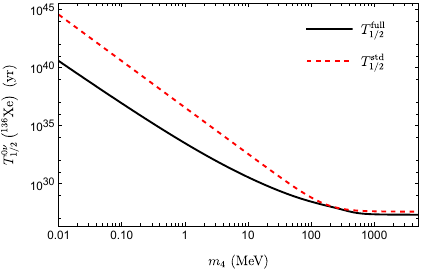}
\caption{Predicted \znubb lifetimes for a scenario with one BSM neutrino of mass $m_4$. Adapted from ref.~\cite{Dekens:2024hlz}.}
\label{fig:3p1}
\end{minipage}
\end{figure}

Fig.~\ref{fig:amps} shows the \znubb amplitudes for a neutrino in the intermediate mass range. In black (solid) is the total amplitude, with the blue (dashed) and red lines showing the hard and potential pieces, $\amp^\text{(h)}$ and $\amp^\text{(p)}$ respectively. The black dotted line shows the amplitude obtained using eq.~\eqref{eq:stdformula} for comparison. The difference in predicted lifetimes when using the EFT computation (in black) for amplitudes, compared to the standard prescription (in red, dashed), is shown for a simple (however, unrealistic) model with one HNL in fig.~\ref{fig:3p1}.

\section{The $3+2$ scenario}

We now turn to a more realistic model with two RHNs. This results in five Majorana neutrinos, with the lightest neutrino being massless. The neutrino mass matrix takes the form
\begin{align}
M_\nu = \left(\begin{matrix}0 & m_D \\ m_D^T & M_M\end{matrix}\right)\,,
\end{align}
where $m_D$ stands for the Dirac masses, while the RHN mass matrix is given by
\begin{align}
M_M = \left(\begin{matrix}\M\left(1 - \frac{\mu}{2}\right) & 0 \\ 0 & \M\left(1 + \frac{\mu}{2}\right)\end{matrix}\right)\,,
\end{align}
for the average Majorana mass $\M$ and a small degeneracy-breaking parameter $\mu$.

The top-left block of the $5\times5$ mixing matrix will be given by the PMNS matrix. We denote the $3\times 2$ mixing matrix of the HNLs with the SM interaction eigenstates with $\theta \simeq m_D M_M^{-1}$; i.e., $\theta_{\alpha i}$ stands for the mixing of the neutrino flavour $\alpha$ with the $i^\text{th}$ mass eigenstate. We also define here $U_{\alpha i}^2 = |\theta_{\alpha i}|^2$, $U_\alpha^2 = \sum_{i}U_{\alpha i}^2$.

The relevant combination for \znubb is~\cite{deVries:2024rfh}
\begin{align}
\label{eq:aeff}
\amp_\text{eff} \equiv &\sum_{i=1}^5 \mathcal{U}_{ei}^2 m_i \amp(m_i)\nonumber\\
\Rightarrow|\amp_\text{eff}|\simeq & \left|\sum_{i=1}^3 m_i \mathcal{U}_{ei}^2 \left(\amp(0) - \amp(\M)\right) + e^{i\lambda}\mu U_e^2\frac{\M^2}{2}\amp'(\M)\right|\,,
\end{align}
where $\lambda$ is a completely free phase. As a result, HNLs can either enhance or reduce the \znubb decay rate, depending on the value of their masses, their mixing with SM neutrinos, and the phase $\lambda$. 

Given that there has been no observation \znubbns, this relation already puts an upper limit on $\mu U_e^2$. $U_e^2$ is an experimentally relevant parameter for colliders and other searches. It can also be constrained from below from cosmological considerations such as Big Bang Nucleosynthesis (BBN) and the seesaw limit. Finally, if we also impose that the entire matter-antimatter asymmetry is explained within this framework via leptogenesis, more chunks of the parameter space can be excluded (see also refs.~\cite{Hernandez:2016kel,Drewes:2016lqo} for earlier studies combining \znubb and leptogenesis).

\subsection{The future}

The fact that the next generation of \znubb experiments can probe the entirety of the standard 3 light neutrino band in the inverted hierarchy points to immense potential to extract information on the neutrino sector. Assuming IO, if no signal is observed in the next generation, it would mean that the second piece in eq.~\eqref{eq:aeff} cancels either partially or fully against the light neutrino contribution, giving us stringent lower limits on $U_e^2$, depending on the mass splitting $\mu$. Consequently, combined with the requirement of leptogenesis, the viable parameter space gets bounded from all directions, and the remaining space will be readily testable at future experiments, such as DUNE, SHiP, and the high-luminosity phase of LHC.

\begin{figure}[h!]
\centering
\includegraphics[width=0.7\linewidth]{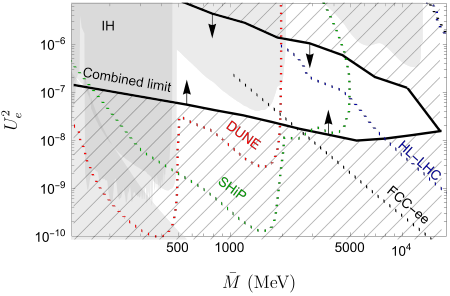}
\caption{Future prospects for the $3+2$ model in the inverted mass ordering of neutrinos. With two orders of magnitude improvement in \znubb limits, the hatched region can be ruled out if the model is to explain the entire BAU. The current limits from experiments and cosmology are shown in grey. Future experimental programmes, shown with dotted lines, will be able to probe all of the remaining allowed parameter space. Taken from ref.~\cite{deVries:2024rfh}.}
\label{fig:exclusion}
\end{figure}

In fig.~\ref{fig:exclusion} we show the $\M-U_e^2$ plane with the current experimental limits and cosmological bounds in grey~\cite{Bondarenko:2021cpc}, and the regions excluded with a combined limit from future non-observation of \znubb (assuming two orders of magnitude improvement over the current limit for ${}^{136}$Xe) and the requirement of correct BAU hatched out. Future searches are shown with dotted lines~\cite{SHiP:2018xqw,Gunther:2023vmz,Drewes:2019fou,Blondel:2022qqo}.

The requirement of leptogenesis also limits the value of $\mu$. We find that the unconstrained region, again assuming IO and no \znubb signal in the near future, will force $\mu$ to be large enough that it can be resolved at experiments. As a result, the $3+2$ model will be completely tested in the near future for IO, while for normal mass ordering the limits from the next generation of experiments might still fall short of probing the entire allowed space.

\section{Conclusions}

Neutrinoless double beta decay can potentially confirm the Majorana nature of neutrinos. We revisit the calculation of rates of this LNV process from a chiral EFT perspective, and show that for light-to-intermediate mass of neutrinos, the amplitudes can differ significantly from the usual functional form considered in the literature. Given the potential of \znubb to probe BSM physics, we study a minimal setup with two right-handed neutrinos that can explain neutrino masses and the matter-antimatter asymmetry simultaneously. It is seen that for the inverted neutrino mass ordering, the experimental prospects look encouraging as the next generation will be able to probe the entire allowed parameter space in this $3+2$ model. For normal mass ordering, however, complete testability of the model would require further experimental progress.

\acknowledgments

I thank Wouter Dekens, Jordy de Vries, Daniel Castillo, Javier Men\'endez, Emanuele Mereghetti, Pablo Soriano, Guanghui Zhou, Marco Drewes, Yannis Georis, and Juraj Klari\'c for collaboration on works presented here. I'd also like to thank the organisers of the workshop for making it a pleasant experience.

\bibliographystyle{JHEP}
\bibliography{refs}



\end{document}